# Clasificarea distribuită a mesajelor de e-mail

Florin Pop
florinpop@cs.pub.ro

Diana Petrescu
diana.petrescu@gmail.com

Ştefan Trauşan-Matu
trausan@racai.ro

Facultatea de Automatica şi Calculatoare, Universitatea "Politehnica" din Bucureşti

Splaiul Independenţei, nr.313, Bucureşti, Sector 6, 060042 România

**REZUMAT**

O componentă de bază în cadrul aplicaţiilor internet o constituie poşta electronică şi diversele sale implicaţii. Lucrarea propune un mecanism de clasificare automată a mesajelor de poşta electronică şi de creare dinamică a categoriilor la care aparţin aceste mesaje. Mecanismele propuse vor avea la bază tehnici de prelucrarea limbajului natural şi vor fi proiectate pentru a uşura interacţiunea om-maşina în aceasta direcţie.

**Categorii şi descriptori ai subiectelor**

I.7 Document and text processing.

**Termeni generali**

*Algorithms, Documentation, Experimentation, Human Factors.*

**Cuvinte-cheie**

Poşta electronică, clasificare, prelucrarea limbajului natural, calcul distribuit.

## 1. INTRODUCERE

Una dintre acţiunile cotidiene pentru orice persoană, din orice domeniu este citirea poştei electronice (e-mail). Mesajele trimise pot fi extrem de numeroase şi de mai multe feluri, fiind necesar un mecanism de clasificare a acestora pe categorii. Clienţii actuali de (programele care furnizează accesul la) e-mail au posibilitatea de a crea directoare şi reguli speciale care se transpun în mecanisme de filtrare a mesajelor primite. Prelucrarea automata şi clasificarea făcută pe baza unei astfel de prelucrări poate duce la o organizare adecvată a mesajelor.

Ne propunem descrierea unui mecanism de clasificare automată a mesajelor primite prin poşta electronica şi de creare a categoriilor cărora pot aparţine aceste mesaje. Am imaginat o arhitectura bazată pe grafuri bipartite în care o mulţime reţine şi modelează mesajele de e-mail, iar cealaltă mulţime conţine categoriile de mesaje. Un mesaj poate aparţine mai multor domenii în această arhitectură. Mecanismul trebuie să conţină un algoritm de decizie care este baza mecanismului de clasificare.

Interacţiunea dintre utilizator şi un client de e-mail dotat cu acest mecanism va fi eficientă deoarece lucrul pe diferite domenii va fi concentrat, având la dispoziţie toate mesajele ce ţin de domeniul respectiv, dar care ar putea fi utile şi unui alt domeniu.

Ne propunem de asemenea şi analiza performanţei sistemului şi posibilitatea includerii acestuia în clienţii de e-mail existenţi.

În secţiunea următoare se face o prezentare a principalilor clienţi de mail si a facilităţilor acestora. În secţiunea a 3-a este propus mecanismul de analiză, de prelucrare şi de stocare a mesajelor de e-mail.

## 2. CLIENŢI DE E-MAIL EXISTENŢI

Poşta electronica, sau pe scurt, e-mail, reprezintă o metodă de a compune, trimite şi primi mesaje pe cale electronică [1]. Termenul de e-mail se foloseşte atât cu referinţă la sistemul de e-mail prin Internet, bazat pe protocolul SMTP (Simple Mail Transport Protocol), cât şi la cel bazat pe comunicaţia prin Intranet.

E-mail-ul datează încă din 1965, precedând astfel reţeaua Internet. Sistemul de e-mail s-a extins rapid, permiţând utilizatorilor să transfere mesaje între calculatoare. Printre primele sisteme care au avut aceasta facilitate se numără AUTODIN (1966) şi SAGE [1].

Clienţii de e-mail reprezintă programe de calculator utilizate pentru citirea şi scrierea de mesaje e-mail. Iniţial, aceste programe s-au dorit a fi simple facilităţi pentru citirea mesajelor şi salvarea lor, dar, o data cu dezvoltarea reţelei Internet, clienţii de mail au evoluat ajungând azi sa ofere numeroase facilităţi precum: crearea de filtre personalizate, administrarea mai multor conturi simultan, administrarea de grupuri, signaturi personalizate, etc.

Asemenea protocoalelor de comunicaţie cu serverul de mail, cei mai mulţi clienţi de e-mail respectă protocoalele POP3 (Post Office Protocol version 3) şi IMAP (Internet Message Access Protocol). IMAP este folosit în special pentru stocarea e-mail-urilor pe server, în timp ce POP3 este folosit pentru descărcarea mesajelor la client. Protocolul pentru trimiterea mesajelor este SMTP.

Un alt standard important care este suportat de marea majoritate a clienţilor de e-mail este MIME (Multiporpose

Internet Mail Extension) folosit pentru trimiterea fișierelor atașate mesajului propriu-zis.

In prezent exista o larga varietate de clienți de mail:
- clienți cu interfața grafică: Microsoft Outlook Express, Microsoft Outlook, Mozilla Thunderbird, Mozilla Mail & Newsgroups, Netscape, Opera M2, Lotus Notes, Appel Mail, etc.
- clienți în mod text: Elmo, Gnus, Pine, etc.
- clienți de e-mail pentru Web: Cidadel, WebPine, SquirrelMail, Zimbra, etc.

Mulți dintre acești clienți de e-mail au filtre statice extrem de performante. Aceste filtre se folosesc atât pentru identificarea mesajelor tip „spam" (mesaje nedorite) cat și pentru organizarea emailurilor pe directoare în funcție de anumite categorii. Având în vedere faptul ca aceste filtre sunt automate, uneori apar erori, de exemplu mesaje care nu sunt spam sunt marcate ca spam. Rata erorilor este însă foarte scăzuta. Nici un filtru automat nu este perfect [2].

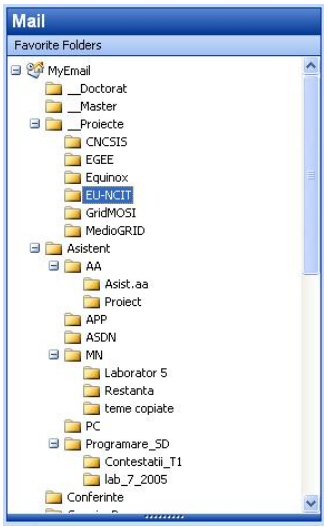

Figura 1: Un exemplu de organizare a mesajelor

În Figura 1 este prezentat un exemplu de organizare a mesajelor cu ajutorul filtrelor statice. Au fost create categorii pe trei nivele. Există categorii principale și sub-categorii ale acestora.

Anumiți clienți de e-mail folosesc o metodă de filtrare numită filtrare Bayesiană [8] Aceste filtre sunt capabile să învețe ca urmare a acțiunilor utilizatorului. Ele pornesc de la un set de reguli inițiale, iar pe măsură ce utilizatorul identifică erori în rezultatul filtrării, își actualizează setul de reguli. Filtrele Bayesiene se folosesc în special pentru filtrarea mesajelor de tip spam, rata erorii fiind mică datorită faptului că învață pe măsură ce utilizatorul marchează mesajele ca spam.

Nu toți clienții de e-mail beneficiază de aceasta facilitate. Pentru cei care nu o au, există tehnologia SpamSieve, care creează aceste filtre Bayesiene în clientul de e-mail.

In continuare vom prezenta câțiva clienți de e-mail existenți, posibilitățile și limitările lor în ceea ce privește filtrarea automata a mesajelor.

## 2.1 Mozilla Thunderbird

Mozilla Thunderbird este un client de e-mail care rulează pe sistemele de operare Windows, Mac OS X, Linux, BSD și Unix. Suportă protocoale de comunicație POP3, IMAP4, SMTP, NNTP și SMTP/Auth.

Thunerbird poate administra mai multe conturi de e-mail și de știri. De asemenea, pune la dispoziție facilități precum: căutarea rapida a unui mesaj după diferite categorii – subiectul mesajului, cel care la trimis, destinatarul principal și cei secundari, etc., salvarea mesajelor în diferite directoare virtuale, filtrare de mesaje și etichetarea mesajelor.

Pentru filtrarea mesajelor se folosește un filtru bayesian, în combinație cu o listă de adrese permise. Utilizatorul își poate crea propriile filtre pentru fiecare cont de mail. Pentru crearea unui filtru este necesar ca utilizatorul să stabilească regulile după care se face filtrarea – numele celui care trimite mesajul sau subiectul acestuia, spre exemplu, și să decidă ce acțiune se va efectua asupra mesajului – va fi mutat într-un director anume, va fi șters, etc. Aceste acțiuni se fac manual, iar programul nu este capabil sa învețe din corecțiile utilizatorului [3].

## 2.2 Outlook Express

Acest client rulează numai pe sistemul de operare Windows. Este unul dintre cei mai populari clienți de e-mail deoarece se distribuie împreună cu sistemul de operare Windows, însă este recunoscut și ca unul dintre cei mai vulnerabil clienți în fața virusurilor și a "viermilor" [1]. Respectă protocoale de comunicație precum POP3, SMTP, IMAP4.

Outlook Express poate administra mai multe conturi de e-mail și de știri și permite crearea de filtre pentru fiecare cont în parte. Similar ca și la Thunderbird, utilizatorul trebuie să își creeze propriile reguli și acțiuni pentru mesajele pe care le dorește filtrate.

O deficiență majoră a acestui client este lipsa unui filtru de spam eficient. Alternativa o reprezintă blocarea adresei celui care trimite e-mail-ul identificat de utilizator ca fiind spam [4].

## 2.3 Eudora 6

Eudora rulează pe sistemele de operare Windows și Mac OS X. Respectă protocoalele POP3, SMTP și IMAP.

Clientul de e-mail Eudora are un filtru bayesian încorporat. Pentru a fi eficient, utilizatorul marchează mesajele spam, iar pe baza acestui marcaj, Eudora își construiește regulile pe baza cărora se decide dacă mesajele sunt tratate ca spam sau nu.

Eudora permite organizarea mesajelor în directoare și crearea unor interogări complexe pentru identificarea rapidă a unui anumit mesaj.

Un dezavantaj al acestui client este faptul ca nu permite stocarea mesajelor trimise și pe server (așa cum ar trebui conform IMAP). Astfel, mesajele trimise nu pot fi accesate decât de pe calculatorul pe care au fost primite [5].

## 2.3 Pine

Pine este un client de e-mail folosit pe sistemele de operare Windows, Mac OS X și Unix. Versiunea pentru Unix este tip linie de comandă. – Pine este un client de e-mail mod text. Pine acceptă protocoalele de comunicație POP3, SMTP, IMAP4, NNTP (Network News Transfer Protocol) și LDAP (Lightweight Directory Access Protocol).

Acest client este capabil sa administreze mai multe conturi de e-mail și permite utilizatorului să își definească reguli de filtrare pe mesaje. Aceste reguli se scriu într-un fișier de configurație. Întrucât este un client mod text, editarea regulilor este relativ dificilă pentru un utilizator neexperimentat [6].

## 2.4 SquirrelMail

SquirrelMail este un client de e-mail pentru browsere Web. Avantajul sau este dat de faptul ca poate fi instalat pe orice server Web care are PHP4 și are acces la un server IMAP și SMTP.

Acest client permite utilizatorului să își citească mesajele de pe orice calculator prin intermediul unui browser Web ușor de folosit. Mesajele pot fi organizate în directoare, se pot crea filtre pentru sortarea automata a mesajelor. În ceea ce privește filtrele pentru spam, acestea se bazează pe o listă neagra de adrese – mesajele de la aceste adrese fiind considerate spam.

Un dezavantaj al acestui client este faptul că nu păstrează conexiunile IMAP. Acesta provine din faptul ca este realizat în PHP [7].

Aceste aspecte privitoare la metodele de prelucrare a mesajelor de poștă electronică duc la o direcție în cadrul interacțiunii om-mașina, direcție în care se pot dezvolta diverse funcționalități în funcție de utilizatori. Un exemplu care trebuie menționat ar fi acela al folosirii acestui serviciu de poștă electronică de către persoanele cu dizabilități de vedere. Pentru aceștia, categoriile de mesaje și crearea lor automată ar însemna un proces de auto-organizare menit să înlesnească accesul la informația primită.

## 3. MODELUL ARHITECTURAL PROPUS

Din analiza clienților de e-mail și mai ales a funcționalității acestora, propunem un model ce poate fi integrat într-un astfel de client și care poate fi configurat de către utilizatori. Vom prezenta un model al metodei de analiză și un model funcțional.

### 3.1 Modelul intern

Modelul ce corespunde cu descrierea făcută în introducere este acela al unui graf bipartit (vezi Figura 2). Cele două mulțimi ale acestui graf descriu cele două entități: mesajul de poștă electronică și categoriile.

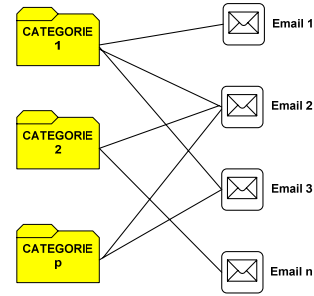

Figura 2. Modelul intern

Prima mulțime, cea a mesajelor de e-mail, poate fi considerată în primă formă chiar mulțimea mesajelor în formă brută venite de la serverul de e-mail. Se poate adăuga un rezumat al acestuia și un număr de cuvinte cheie care pot descrie cât mai bine conținutul mesajelor. Pe baza acestora se poate construi mecanismul de clasificare. Ca rezultat, vom avea legăturile descrise în figura 2. Tehnicile de analiză și prelucrare a conținutului mesajelor pot fi diverse metode specifice prelucrării limbajului natural. O categorie care poate fi luată în seamă este cea a metodelor de minerit de date și de prelucrare statistică.

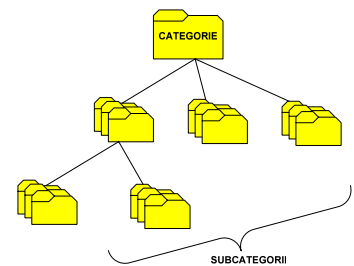

Figura 3. Analiza în detaliu: sub-categorii

Mulțimea categoriilor este rezultatul procesului de clasificare. Inițial putem avea o mulțime vidă a categoriilor sau o mulțime creată de utilizator. În funcție de rezultatele mecanismului de clasificare se pot crea noi categorii. Fiecare mesaj va trebui să aparțină uneia sau mai multor categorii. Fiecare categorie va conține cel puțin un mesaj. Pentru fiecare categorie în parte se poate crea un mecanism specific de clasificare, ce va avea ca rezultat crearea de sub-categorii (vezi figura 3). Rolul acestor sub-categorii este de a oferi o rezoluție cât mai bună asupra informațiilor dintr-o categorii.

Mecanismul de clasificare poate fi conceput ca un mecanism distribuit în care pe mai multe mașini pot exista mai mult servere de e-mail ce pot stoca mesajele primite de un utilizator (de exemplu un utilizator cu mai multe adrese de e-mail stocate pe servere diferite). Utilizatorul va percepe și va avea acces la informația conținută de mesaje din orice locație cu ajutorul clientului de e-mail care îi poate oferi această facilitate.

Componenta distribuită a mecanismului va conduce la stocarea mesajelor doar pe server, categoriile conținând descrierea și locația unui mesaj.

### 3.2 Mecanismul de comunicație

Accesul la mesajele de mail care sunt stocate pe server poate fi făcut prin intermediul protocoalelor cunoscute[1]. Mecanismul de comunicație este descris în Figura 4.

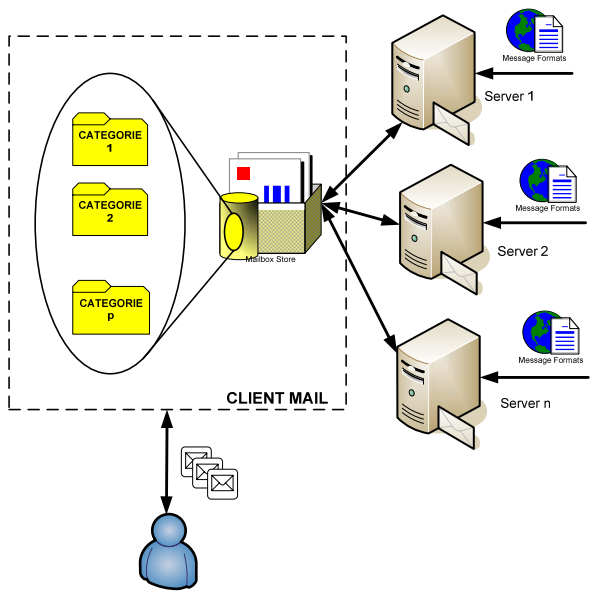

Figura 4. Mecanismul de comunicație

Interacțiunea om-calculator este evidențiată în acest model de comunicație doar la nivelul clientului. Este astfel pus la dispoziție un mecanism ergonomic care nu necesită cunoștințe avansate de protocoale de comunicație sau de calcul distribuit. Folosind mecanismele de calcul distribuit și implementând mecanismul la nivel de client, putem identifica o componentă centralizată, la nivel de client (în care toate categoriile trebuie sa fie vizibile) susținută de partea de calcul distribuit in procesul clasificării.

---

[1] Nu considerăm necesară modificarea acestor protocoale (cum ar fi SMTP, POP3 sau alte protocoale), toate acțiunile care se vor face în procesul de clasificare fiind implementate la nivelul clienților de mail.

Întreg procesul de comunicație din Figura 4 se suprapune peste protocoale existente. Utilizatorul va avea la dispoziție toate mesajele corespunzătoare unei categorii. Se pot, de asemenea imagina diverse posibilități de configurare a categoriilor și de stocare a informațiilor corespunzătoare (baze de date, fișiere XML, etc.).

### 4. CONCLUZII

Am pornit de la analiza clienților de e-mail care sunt aplicații cu diverse implicații în domeniul interacțiunii om-mașină, analizând funcționalitățile acestora. Bazându-ne pe acestea, am descris un model de clasificare a mesajelor de e-mail lăsând ca dezvoltări ulterioare implementarea și testarea sistemului, precum și compararea acestuia cu diverse sisteme existente. Se va elabora un model de evaluare care va ține cont de diversele specificații ale utilizatorilor.

Implementarea mecanismelor poate aduce componente noi clienților de e-mail fără modificarea funcționalității acestora. Analiza mesajelor poate conduce la folosirea de noi tehnici de semantică web și de prelucrare a informațiilor, toate acestea aducând un element de noutate în interacțiunea om-calculator pe partea de poștă electronică.

### 5. REFERINȚE